\documentclass{article}
\usepackage{spconf,amsmath,graphicx}
\usepackage{etoolbox,siunitx}
\usepackage{multirow}
\usepackage{etoolbox,siunitx}
\robustify\bfseries
\usepackage{booktabs}
\usepackage{balance}
\usepackage{amssymb}
\usepackage{bm}
\usepackage{arydshln}
\usepackage{caption} 
\usepackage{xcolor}
\usepackage{enumitem, setspace, hyperref}

\input{def.set}

\title{SpaIn-Net: Spatially-Informed Stereophonic Music Source Separation }
\name{Darius Petermann and Minje Kim}
\address{Indiana University, Department of Intelligent Systems Engineering, Bloomington, IN, USA 47408}

\begin{document}
\ninept
\maketitle
\begin{abstract}

With the recent advancements of data driven approaches using deep neural networks, 
music source separation has been formulated as an instrument-specific supervised problem. While existing deep learning models implicitly absorb the spatial information conveyed by the multi-channel input signals, we argue that a more explicit and active use of spatial information could not only improve the separation process but also provide an entry-point for many user-interaction based tools. To this end, we introduce a control method based on the stereophonic location of the sources of interest, expressed as the panning angle. We present various conditioning mechanisms, including the use of raw angle and its derived feature representations, and show that spatial information helps.  Our proposed approaches improve the separation performance compared to location agnostic architectures by 1.8 dB SI-SDR in our Slakh-based simulated experiments. Furthermore, the proposed methods allow for the disentanglement of same-class instruments, for example, in mixtures containing two guitar tracks. Finally, we also demonstrate that our approach is robust to incorrect source panning information, which can be incurred by our proposed user interaction. 
\end{abstract}
\section{Introduction}
\label{sec:intro}

Musical source separation (MSS), a task consisting in isolating various musical constituents from a given music mixture, has been an active research area for decades now. The problem is challenging due to the typical \textit{underdetermined} nature of musical signals (i.e., lesser number of channels than sources), hence it has been addressed via machine learning, e.g., spectrogram decomposition \cite{GilletO2008ieeeaslp,  OnoN2008eusipco}. 
Recently, deep learning and data driven approaches have advanced this field of study significantly. A typical deep learning-based MSS system can be trained in a supervised fashion by comparing the model's output to the ground-truth source signals. It is also common to employ the concept of masking in the feature space, such as ideal ratio masking (IRM) \cite{NarayananA2013icassp} on the coefficients of the short-time Fourier transform (STFT) \cite{HuangP2014ismir}, while a direct waveform estimation is also common, such as seen in Wave-U-Net or Demucs \cite{Lluis2019, StollerD2018waveunet, DefossezA2021demucs}. 

\begin{figure}[t]
    \centering
        \includegraphics[scale=0.5]{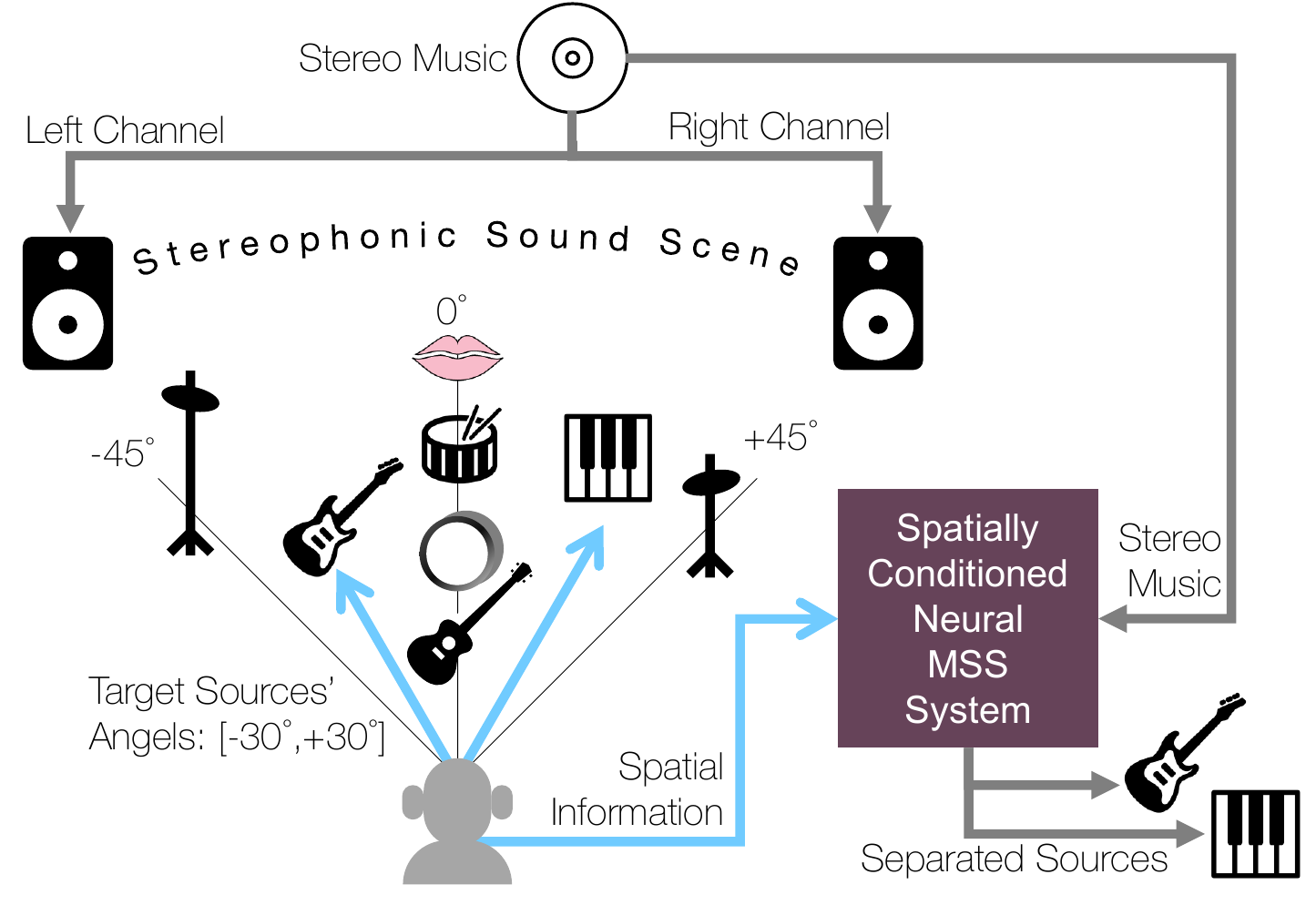}\vspace{-.1cm}
    \caption{Diagram of the overall proposed system. Notice that the stereo field location is addressed in degrees.}
    \label{fig:panning_scheme}
\end{figure}

In this paper, we focus on the stereophonic mixtures. In music especially, stereo channel settings are a widely popular format and usually preferred over monophonic mixtures, since it conveys a larger spatial field for a more enjoyable listening experience. Discussing the professional stereophonic mixing process is out of the scope of this paper as it is artistic and complicated. It is however important to note that each music source tends to have unique stereophonic characteristics, such as a panning location in the stereophonic panorama. For example, Fig. \ref{fig:panning_scheme}  portrays what a typical panning configuration for Pop music could look like. These typical configurations can however change depending on the music genre, instrumentation, and the mixing engineer's creative freedom over the process, making supervised learning challenging.

Indeed, stereophonic MSS has added another dimension to the MSS problem. There are models assuming a source-specific spatial panning position and disjoint orthogonality among sources in the time-frequency domain, such as DUET \cite{RickardS2007duet}, ADRess \cite{BarryD2004adress, SofianosS2010icassp}, and PROJECT \cite{FitzgeraldD2016project}. While these models are strictly instrument-agnostic, 
it is known that the source and spatial modeling approaches can be combined together as in multichannel nonnegative matrix factorization \cite{OzerovA2010multichannel-nmf} and the separation of the main-versus-accompaniment using the source-filter model \cite{DurrieuJ2009eusipco}. Likewise, knowing of or assuming about the source locations in the stereophonic audio scene can help improve source separation. 

Here, we claim that MSS could further benefit from some additional, high-level, spatial information, if it is provided more directly. In that regard, our approach can be seen as a variant of \textit{informed} source separation \cite{LiutkusA2013informed}, where aligned scores \cite{DuanZ2011soundprism}, the user's query \cite{LeeJH2019query}, and even the user's scribble on the spectrograms \cite{BryanN2013icml} can serve as the auxiliary information. Similarly, we envision that the sources' spatial locations can be used as the auxiliary input to a machine learning-based MSS system as shown in \cite{ChenZ2018guided} for speech separation.
We postulate that spatial information would be useful when the other features, such as spectral, timbral, or temporal characteristics, are not discriminative enough, e.g., in unseen instruments or artificially synthesized sound.

To this end, we propose to \textit{condition} a deep neural network (DNN) using the spatial information of the sources of interest, which we call spatially informed networks (SpaIn-Net). Injecting prior-knowledge into deep learning has been well investigated for MSS applications, for example, the target source's label \cite{SeetharamanP2019class, Meseguer-BrocalG2019conditioned-u-net}, a query audio signal that describes the target source \cite{LeeJH2019query}, etc. To the best of our knowledge, the proposed model is the first attempt in the informed MSS literature to condition a DNN using spatial information of the sources. 

The proposed model applies the conditioning idea to one of the state-of-the-art MSS systems, called Open-Unmix + CrossNet (XUMX) \cite{SawataR2020cumx}.
We investigate various conditioning mechanisms and show that they overall improve the MSS performance compared to the baseline unconditioned XUMX model. 
Note  that the system  also adds an interactive interface entry point, allowing for an \textit{inaccurate} user input that still helps MSS, opening up a new direction to user-centered applications. The robustness to the noisy user input differentiates SpaIn-Net from the setup in \cite{ChenZ2018guided}.

\section{Methodology}
\label{sec:method}

\subsection{Baseline Model}

Our baseline model, the XUMX architecture, was introduced as part of the \textit{Music Demixing Challenge 2021} \cite{MitsufujiY2021music} as an extension of Open-Unmix (UMX) \cite{StoterFR2019open-unmix}. The XUMX model's superiority comes from its advanced loss functions. First, the \textit{multi-domain} loss function computes the source reconstruction loss both in the frequency and time domain, for the former mean-squared loss compares the magnitudes of source and reconstruction, while the latter employs weighted signal-to-distortion ratio (wSDR) on the time-domain signals, directly. Second, the model also employs a \textit{combination} loss that examines all partial mixtures and their reconstruction, e.g., the mixture of guitar and bass versus the mixture of the estimated guitar and bass, and so on. In this work, we opt to use of the multi-domain loss as the sole loss function.
The concept of combining sources is also used within the model where the source-specific features are averaged up across the original UMX network's source-specific extraction streams. We inherit the XUMX model to construct our baseline and the proposed systems, although we opted out of the combination loss which degrades the separation performance in our same-source separation task. 

\subsection{Spatial embeddings}

Since the conditioning process combines heterogeneous data types, which in our case consist of stereo audio signals and the sources' spatial information, it needs a careful design to benefit from both modalities. 
First, it is reasonable to assume that the audio signals are in high dimensional space. In our XUMX baseline, for example, the input signal goes through STFT, resulting in an $F$ dimensional input vector at $t$-th time step, where $F$ is defined by the frame size. Meanwhile, as for the spatial conditions, we opt to use the angle of the source instrument's panning location in the stereophonic sound field as illustrated in Fig. \ref{fig:panning_scheme}. For example, if the user wants to separate guitar and piano, the corresponding panning location will be $-30^\circ$ and $+30^\circ$, respectively. These scalars are obviously not descriptive enough when it comes to professionally engineered music, where the instruments can have ambient effects that disperse the perceived panning location of the source.
However, considering the potential user interface that may benefit from its simplicity, we employ the scalar angle value to inform the MSS system. 

One obvious approach to combine these two types of information is to concatenate the angle value to the spectrum, e.g., by appending each of the $K$ angle values of $K$ sources to each of the corresponding XUMX source-specific inference streams, forming an $F+1$ dimensional vector per inference stream. 

While appending the scalar to the input vector might be a valid way, we investigate more elaborated methods to carefully examine the impact of spatial information on MSS. 
We observe that the main issue might be that the two dimensions are very different, e.g., $F\gg 1$. 
Out of various other ways to condition a neural network, such as FiLM \cite{PerezE2018film}, in this paper, we adopted the positional encoding method proposed in the Transformer model \cite{VaswaniA2017transformer} that expands the scalar variable's dimension using sinusoids. The original positional encoding scheme converts a nonnegative integer value (e.g., the word order index within the input sentence) into a sinusoidal function, represented in a $D$-dimensional vector. The shape of the output positional embedding vector differs based on the scalar input for discrimination. However, the original formulation is defined only for nonnegative integers, thus necessitating a variant to cover negative numbers, i.e., source positions on the left channel.

\begin{figure}[t]
    \centering
        \includegraphics[scale=0.22]{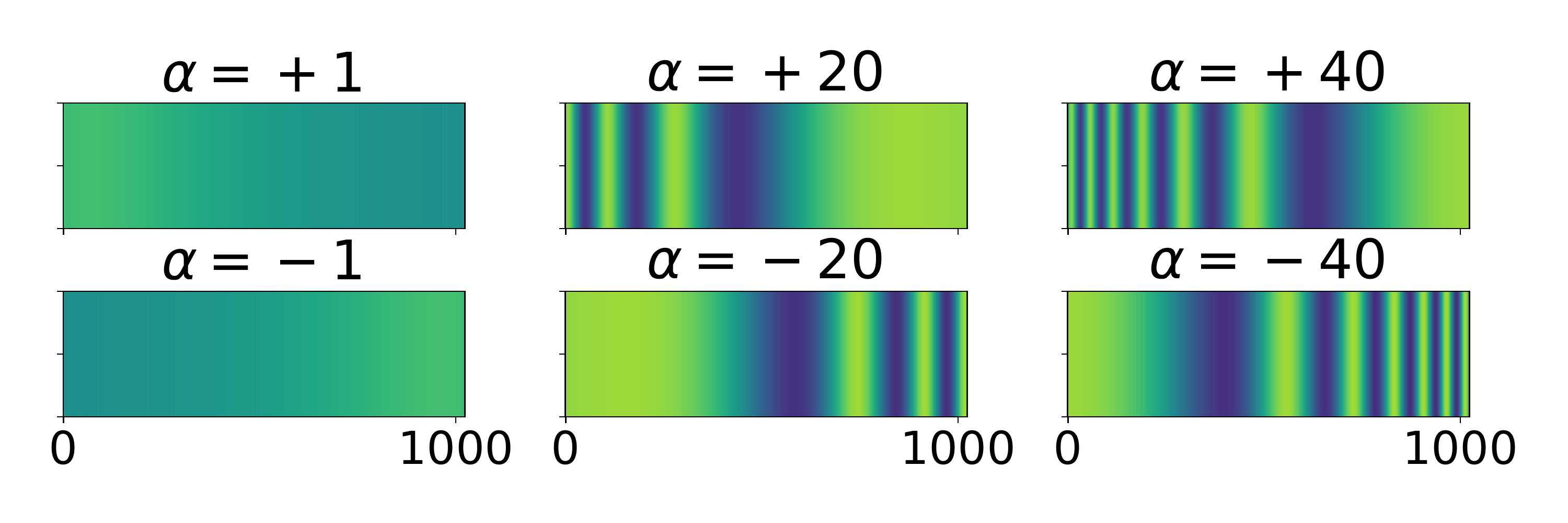}\setlength{\textfloatsep}{2pt}
    \caption{Examples of positional encoding embeddings for various angular degree values. Note that the further apart from the center the values are, the more disparity is reflected in their resulting positional encoding maps.}
    \label{fig:positional_encoding}
\end{figure}

Hence, our proposed positional encoding is designed to create the vector version of both positive and negative scalars. First, the positive side is defined similarly to the Transformer's. $\calP$ is a function of the angle value in degree $0\leq\alpha\leq +45$ and the dimension index $i$ that varies from $0$ to $D/2$, where $D$ is the target dimension:
\begin{equation}
\calP(2i, \alpha)=\sin \left(\frac{\alpha}{45^{\frac{2 i}{D}}}\right), \quad \calP(2i+1, \alpha)=\cos \left(\frac{\alpha}{45^{\frac{2 i}{D}}}\right).
\end{equation}
Here, $\calP$ is defined by alternating sine and cosine functions. For a given input scalar $\alpha$, the sinusoidal function ``slows down" its frequency exponentially as the dimension $i$ increases. The result is a sinusoidal function that gradually decreases its frequency in the higher dimension (Fig \ref{fig:positional_encoding}, the first row). $\alpha$ contributes to the overall frequency of this resulting sinusoidal function: the smaller $\alpha$ is, the more it reduces the overall frequency and vice versa. For the negative angles, we flip these sinusoids in the left-right direction, so that the ripple area (defined here as the faster changing frequency portion of the positional encoding vector) appears on the opposite side:
\begin{equation}
\calN(2i, \alpha)\!=\!\sin \left(\frac{\alpha}{45^{\frac{D-2 i}{D}}\!}\right),~\calN(2i + 1,\!\alpha)\!=\!\cos \left(\frac{\alpha}{45^{\frac{D-2 i}{D}}\!}\right).
\end{equation}
Note that in Fig \ref{fig:positional_encoding} $D=1024$ is a dimension chosen empirically among other options (e.g., $512$, $2048$, etc.).



\subsection{Conditioning mechanisms}


We condition the XUMX baseline by combining the spatial information with the spectrum. Let $\ba_t^{(k)}\in\Real^D$  hold the positional embedding representation of the $k$-th target source's panning location at the given time frame $t$, which is the output of the function $\calP$ or $\calN$ depending on the sign of the angle $\alpha$. 
In this paper we limit our discussion to the static source cases, so we drop the time index $t$ from $\ba^{(k)}$. Also, note that $D=1$ denotes the case where we do not apply the positional encoding and just use the raw angle values, directly. 

While $\ba^{(k)}$ denotes the ``ground-truth" panning angle, we also take the incorrect user input into account. To that end, we employ another notation, a noise-injected angle $\bar{\ba}^{(k)}={\ba}^{(k)}+\epsilon$, where $\epsilon$ is a random deviation amount sampled from a uniform distribution defined between $[-\delta, \delta]$: $\epsilon\sim\calU(-\delta, \delta)$. We will revisit the difference between $\ba^{(k)}$ and $\bar{\ba}^{(k)}$ in the experiments.

Meanwhile, the input mixture signal goes through the first feature extraction step, which is STFT in our XUMX setup. The left and right channel signals go through STFT individually, resulting in a stacked magnitude spectrogram $|\bX|\in\Real_+^{2F\times T}$, whose upper and bottom halves are the left and right channel spectrograms, respectively. Once again, based on the static source assumption, we repeatedly concatenate the spatial embedding $\ba^{(k)}$ to all $T$ spectra. Given that we can have up to $K$ such embedding vectors, the final conditioned input vector at time $t$ is $\left[|\bX_{:,t}|^\top, {\ba^{(k)}}^\top\right]^\top\in\Real^{2F+D}$ for the $k$-th XMUX inference stream, which estimates the $k$-th source. 

Adding the two vectors is also a popular option as in the Transformer model. To that end, the system must make sure that $D=2F$, so that the addition operation holds: $|\bX_{:,t}|+\ba^{(k)}\in\Real^{F}$.



Finally, we also try adaptive instance normalization (AdaIN), which was originally proposed in the context of image style-transfer \cite{HuangX2017adain} with the aim to statistically align a given set of content feature to some target style feature. 
In our case, the style and feature contents denote two different modalities: the spatial information as the content feature $\ba^{(k)}$ and the spectra as the style feature $|\bX_{:,t}|$. AdaIN's goal is to align their mean and standard deviation as follows:
\begin{equation}
\textrm{AdaIN}(|\bX_{:,t}|,\ba^{(k)})= \sigma(\ba^{(k)})\left(\frac{|\bX_{:,t}|-\mu(|\bX_{:,t}|)}{\sigma(|\bX_{:,t}|)}\right)+\mu(\ba^{k}),
\end{equation}
Here, for every frame $t$ we align the mean and variance of the positional encoding $\ba^{k}$ with those of the input spectrogram $|\bX_{:,t}|$.

\section{Dataset and experimental setup}
\label{sec:dataset}

\subsection{The dataset}

Since we seek supervised MSS, access to the isolated ground-truth sources is necessary during training. In this view, we opt to work with the Slakh dataset \cite{ManilowE2019slakh}, which comprises 2,100 songs and 34 instrument categories, for a total of 145 hours of audio data in mono format at a sampling rate of 44.1kHz. Slakh allows a full control of the originally monophonic sources---we freely relocate their stereophonic panning locations using constant power panning laws (See eq.~\eqref{eq:cpp}). Compared to other alternative choices, such as MUSDB \cite{musdb18} or MedleyDB \cite{BittnerR2014MedleyDB}, the use of Slakh avoids unnecessary downmixing of stereophonic original sources, which then have to be upmixed for stereo panning. The downside is that Slakh's stem tracks are originally MIDI sources rendered by virtual instruments. It is also true that our constant power panning may not represent the real-world professional mixing process.
We follow Slakh's original split schemes. However, due to the four-source separation setup which omits other source categories (Sec. \ref{sec:experiment}), the size of each of the training sets naturally reduces to 120 hours. 



\subsection{Mixing procedure}

We use constant power panning (CPP) laws to assign each of the target sources a randomly chosen panning angle.
We ensure that the sources' relative levels in the resulting stereo mix remain the same by using the CPP laws. For an amplitude of the monophonic stem signal $m(n)$ at time index $n$, the CPP law defines the gain of left and right channels as follows:
\begin{equation}
\begin{split}
x_\text{L}(n) &= (\sqrt{2}/2) (\cos{\alpha} + \sin{\alpha}) m(n)\\
x_\text{R}(n) &= (\sqrt{2}/2) (\cos{\alpha} - \sin{\alpha}) m(n),    
\end{split}
\label{eq:cpp}
\end{equation}
which are then multiplied to $m(n)$ to construct the stereo channels $x_\text{L}$ and $x_\text{R}$. The resulting stereo signal $\bx$ should convey a perceived panning location that matches the target angle $\alpha$.

\subsection{The proposed experiments}
\label{sec:experiment}

In order to assess the validity of our approach and showcase that the conditioning spatial information benefits the separation task, we design multiple experimental setups.

\begin{table*}[t]

\centering
  \sisetup{table-format=2.1,round-mode=places,round-precision=1,table-number-alignment = center,detect-weight=true,detect-inline-weight=math}
\caption{BSS Eval improvements observed on \textbf{Task \texttt{4S}} for the CrossNet baseline model and our proposed models: \texttt{4S-D1-CAT-$\alpha_\text{Tr}$} taking the raw angle scalar and \texttt{4S-DF-ADD-$\alpha_\text{Tr}$}. Note that \texttt{4S-D1-CAT-$\bar\alpha_\text{Tr}$} is trained on noisy angle.}
\setlength{\tabcolsep}{6.0pt}
\resizebox{.999\textwidth}{!}{%
\begin{tabular}[t]{lS[table-format=1.1]S[table-format=2.1]S[table-format=1.1]S[table-format=1.1]S[table-format=1.1]S[table-format=1.1]|S[table-format=2.1]S[table-format=1.1]S[table-format=1.1]S[table-format=1.1]S[table-format=1.1]|S[table-format=2.1]S[table-format=1.1]S[table-format=1.1]S[table-format=1.1]S[table-format=1.1]|S[table-format=2.1]S[table-format=1.1]S[table-format=1.1]S[table-format=1.1]S[table-format=1.1]|}
\toprule
\multicolumn{2}{c}{Models}&\multicolumn{5}{c}{\texttt{4S-D0}} &  \multicolumn{5}{c}{\texttt{4S-D1-CAT-$\alpha_\text{Tr}$}} &
\multicolumn{5}{c}{\texttt{4S-D1-CAT-$\bar\alpha_\text{Tr}$}} &\multicolumn{5}{c}{\texttt{4S-DF-ADD-$\alpha_\text{Tr}$}}\\
\cmidrule(lr){3-7} \cmidrule(lr){8-12} \cmidrule(lr){13-17} \cmidrule(lr){18-22}
\multicolumn{2}{c}{Instruments}& {Gtr.} & {Str.} & {Pia.} & {Bas.} & {Avg.} & {Gtr.} & {Str.} & {Pia.} & {Bas.} & {Avg.} & {Gtr.} & {Str.} & {Pia.} & {Bas.} & {Avg.} & {Gtr.} & {Str.} & {Pia.} & {Bas.} & {Avg.}\\
\cmidrule(lr){3-7} \cmidrule(lr){8-12} \cmidrule(lr){13-17} \cmidrule(lr){18-22}
\multicolumn{2}{c}{Mixture SDR}& -12.266 & -22.721 &  -3.511 & -10.172 & -12.167 & -12.266 & -22.721 &  -3.511 & -10.172 & -12.167 & -12.266 & -22.721 &  -3.511 & -10.172 & -12.167& -12.266 & -22.721 &  -3.511 & -10.172 & -12.167\\
\cmidrule(lr){3-22}

&{$\Delta$ SDR } & 10.922 & 15.56 & 8.076 & 9.865 & 11.106 
& \bfseries 12.4  & 17.58 & 8.061 & 11.316 & 12.345 
& 12.168 & 18.063 & \bfseries 8.5  & 11.248 & 12.503 
& 11.983 & \bfseries 18.4  & \bfseries 8.5  & \bfseries 11.7  & \bfseries 12.7  \\

\texttt{4S}&{ISR} 
& 1.462 & 0.174 & 5.627 & 3.557 & 2.705      
& 2.705 & 1.488 & 5.463 & 4.128 & 3.446         
& \bfseries 2.8  & 1.597 & \bfseries 6.0  & 4.052 & \bfseries 3.5      
& 2.313 & \bfseries 1.7  & 5.912 & \bfseries 4.3  & \bfseries 3.5  \\

\texttt{-$\alpha_\text{Te}$}&{SAR} 
& 3.054 & 3.523 & 8.615 & 5.56 & 5.188 
& \bfseries 5.3  & 4.18 & 9.417 & \bfseries 6.8  & 6.416 
& \bfseries 5.3  & \bfseries 4.8  & \bfseries 9.6  & 6.65 & \bfseries 6.6 
& \bfseries 5.3  & 4.736 & 9.461 & \bfseries 6.8  & \bfseries 6.6  \\
&{SIR} 
& 5.348 & 0.511 & 12.623 & 3.983 & 5.616 
& \bfseries 9.0  & 1.091 & \bfseries 14.4  & 7.833 & 8.082 
& 8.733 & \bfseries 2.6  & 14.189 & \bfseries 8.3 & \bfseries 8.4  
& 8.909 & 1.659 & 13.948 & 7.703 & 8.055\\
\cmidrule(lr){3-22}

&{$\Delta$ SDR } 
& 10.922 & 15.56 & 8.076 & 9.865 & 11.106 
& \bfseries 12.299 & 17.528 & 8.042 & 11.264 & 12.283 
& 12.094 & 18.067 & \bfseries 8.492 & 11.217 & \bfseries 12.467 
& 11.657 & \bfseries 18.151 & \bfseries 8.529 & \bfseries 11.68 & \bfseries 12.504 \\

\texttt{4S}&{ISR} 
& 1.462 & 0.174 & 5.627 & 3.557 & 2.705 
& \bfseries 2.685 & 1.517 & 5.43 & 4.082 & 3.428 
& \bfseries 2.733 & 1.604 & \bfseries 5.939 & 3.989 & \bfseries 3.566 
& 2.093 & \bfseries 1.668 & \bfseries 5.937 & \bfseries 4.295 & 3.499 \\

\texttt{-$\bar\alpha_\text{Te}$}&{SAR} 
& 3.054 & 3.523 & 8.615 & 5.56 & 5.188 
& 5.292 & 4.224 & 9.406 & \bfseries 6.682 & 6.401 
& \bfseries 5.407 & 4.789 & \bfseries 9.552 & 6.552 & 6.575 
& 5.27 & \bfseries 5.237 & \bfseries 9.603 & \bfseries 6.72 & \bfseries 6.707 \\

&{SIR} 
& 5.348 & 0.511 & 12.623 & 3.983 & 5.616 
& 8.78 & 0.826 & \bfseries 14.535 & 7.743 & 7.971 
& 8.562 & \bfseries 2.315 & 13.99 & \bfseries 8.326 & \bfseries 8.298 
& \bfseries 9.161 & 1.777 & 14.09 & 7.527 & 8.139\\

\bottomrule
\end{tabular}}
\label{table:results_task1}
\end{table*}
\begin{table*}[t]

\centering
  \sisetup{table-format=2.1,round-mode=places,round-precision=1,table-number-alignment = center,detect-weight=true,detect-inline-weight=math}
\caption{BSS Eval improvements observed on \textbf{Task \texttt{4S2G}} for the CrossNet baseline model and our proposed models: \texttt{4S2G-D1-CAT-$\alpha_\text{Tr}$} taking the raw angle scalar and \texttt{4S2G-DF-ADD-$\alpha_\text{Tr}$}. Note that \texttt{4S2G-D1-CAT-$\bar\alpha_\text{Tr}$} is trained on noisy angle.}
\setlength{\tabcolsep}{6.0pt}
\resizebox{.999\textwidth}{!}{%
\begin{tabular}[t]{lS[table-format=1.1]S[table-format=2.1]S[table-format=1.1]S[table-format=1.1]S[table-format=1.1]S[table-format=1.1]|S[table-format=2.1]S[table-format=1.1]S[table-format=1.1]S[table-format=1.1]S[table-format=1.1]|S[table-format=2.1]S[table-format=1.1]S[table-format=1.1]S[table-format=1.1]S[table-format=1.1]|S[table-format=2.1]S[table-format=1.1]S[table-format=1.1]S[table-format=1.1]S[table-format=1.1]|}
\toprule
\multicolumn{2}{c}{Models}&\multicolumn{5}{c}{\texttt{4S2G-D0}} &  \multicolumn{5}{c}{\texttt{4S2G-D1-CAT-$\alpha_\text{Tr}$}} &
\multicolumn{5}{c}{\texttt{4S2G-D1-CAT-$\bar\alpha_\text{Tr}$}} &\multicolumn{5}{c}{\texttt{4S2G-DF-ADD-$\alpha_\text{Tr}$}}\\\cmidrule(lr){3-7} \cmidrule(lr){8-12} \cmidrule(lr){13-17} \cmidrule(lr){18-22}
\multicolumn{2}{c}{Instruments}& {Gtr1} & {Gtr2} & {Pia.} & {Bas.} & {Avg.} & {Gtr1} & {Gtr2} & {Pia.} & {Bas.} & {Avg.} & {Gtr1} & {Gtr2} & {Pia.} & {Bas.} & {Avg.} & {Gtr1} & {Gtr2} & {Pia.} & {Bas.} & {Avg.}\\
\cmidrule(lr){3-7} \cmidrule(lr){8-12} \cmidrule(lr){13-17} \cmidrule(lr){18-22}
\multicolumn{2}{c}{Mixture SDR}& -15.151 & -16.466 &  -2.346 & -15.019 & -12.245 &-15.151 & -16.466 &  -2.346 & -15.019 & -12.245 & -15.151 & -16.466 &  -2.346 & -15.019 & -12.245 & -15.151 & -16.466 &  -2.346 & -15.019 & -12.245\\
\cmidrule(lr){3-22}

&{$\Delta$ SDR } 
& 9.484 & 10.488 & 7.485 & 12.694 & 10.038 
& \bfseries 12.48 & 13.06 & 7.745 & 14.945 & \bfseries 12.057 
& 12.131 & \bfseries 13.164 & \bfseries 7.832 & \bfseries 15.139 & \bfseries 12.067 
& 12.338 & 12.949 & 7.726 & 13.957 & 11.743 \\

\texttt{4S2G}&{ISR} 
& -1.155 & -0.684 & 6.295 & 2.968 & 1.856 
& \bfseries 1.422 & 1.214 & 6.271 & 3.688 & \bfseries 3.149 
& 1.177 & \bfseries 1.269 & 6.283 & 3.674 & 3.101
& 1.11 & 1.142 & \bfseries 6.393 & \bfseries 3.804 & \bfseries 3.112 \\

\texttt{-$\alpha_\text{Te}$}&{SAR} 
& 5.876 & 5.898 & 9.302 & 5.091 & 6.542 
& 5.858 & 6.214 & \bfseries 9.927 & 6.554 & 7.138 
& \bfseries 6.515 & \bfseries 6.585 & \bfseries 9.876 & 6.581 & \bfseries 7.389 
& 5.743 & \bfseries 6.56 & 9.818 & \bfseries 6.688 & 7.202 \\

&{SIR} 
& -2.251 & -3.218 & 12.942 & 2.055 & 2.382 
& \bfseries 4.781 & \bfseries 4.476 & 15.799 & \bfseries 7.048 & \bfseries 8.026 
& 4.441 & 3.988 & \bfseries 16.101 & \bfseries 6.961 & 7.872 
& 4.039 & 3.207 & 14.024 & 4.864 & 6.533 \\

\cmidrule(lr){3-22}
&{$\Delta$ SDR } 
& 9.484 & 10.488 & 7.485 & 12.694 & 10.038 
& \bfseries 12.096 & 12.929 & 7.704 & 14.82 & \bfseries 11.887 
& 11.73 & \bfseries 13.047 & \bfseries 7.838 & \bfseries 15.147 & \bfseries 11.94 
& 11.963 & 12.666 & 7.721 & 13.693 & 11.511 \\

\texttt{4S2G} &{ISR} 
& -1.155 & -0.684 & 6.295 & 2.968 & 1.856 
& \bfseries 0.885 & 1.27 & 6.243 & 3.559 & 2.989 
& \bfseries 0.871 & \bfseries 1.358 & 6.262 & \bfseries 3.686 & \bfseries 3.044 
& 0.816 & 1.178 & \bfseries 6.373 & 3.639 & \bfseries 3.002 \\

\texttt{-$\bar\alpha_\text{Te}$}&{SAR} 
& 5.876 & 5.898 & 9.302 & 5.091 & 6.542 
& 5.757 & 6.36 & 9.847 & 6.546 & 7.128 
& \bfseries 6.46 & 6.494 & 9.901 & 6.544 & \bfseries 7.35 
& 5.257 & \bfseries 6.837 & \bfseries 9.978 & \bfseries 6.602 & 7.168 \\

&{SIR} 
& -2.251 & -3.218 & 12.942 & 2.055 & 2.382 
& \bfseries 4.808 & \bfseries 3.513 & 15.566 & \bfseries 7.091 & \bfseries 7.745 
& 4.115 & \bfseries 3.547 & \bfseries 16.257 & 6.895 & \bfseries 7.703 
& 3.653 & 2.935 & 14.183 & 4.735 & 6.376 \\

\bottomrule
\end{tabular}}
\label{table:results_task2}
\end{table*}

\begin{table}[t]
\scriptsize
\centering
  \sisetup{table-format=2.1,round-mode=places,round-precision=1,table-number-alignment = center,detect-weight=true,detect-inline-weight=math}
\caption{SI-SDR improvements averaged over all four sources on \textbf{Task \texttt{4S}} for various conditioning approaches. The clean source angles are used for the test signals ($\alpha_\text{Te}$) and the models are trained from accurate source angles (${\alpha}_\text{Tr}$).}\vspace{-.1cm}
\setlength{\tabcolsep}{3.5pt}
{%
\begin{tabular}[t]{l@{\hskip 0.15in} S[table-format=2.1]S[table-format=1.1]S[table-format=1.1]S[table-format=2.1]S[table-format=1.1]}
\toprule
\cmidrule(lr){2-6}
& {\texttt{D0}} & {\texttt{DF$_\text{AdaIN}$}} & {\texttt{D16-CAT}}  & {\texttt{D32-CAT}} & {\texttt{D64-CAT}} \\
\midrule
{Mixture SDR}               & -12.167 & -12.167 & -12.167 & -12.167 & -12.167\\ 
\midrule
{Average SDR}              & 11.106 &  11.6 & 12.307 & 11.729 & 12.476  \\
\bottomrule
\end{tabular}\vspace{-.5cm}
\label{table:results_prelims}
}
\end{table}

\begin{itemize}[noitemsep,topsep=0pt, leftmargin=0in, itemindent=.15in]
    \item \texttt{4S}: The first MSS task involves four distinct musical sources, namely guitar, strings, bass, and piano. 
    \item \texttt{4S2G}: A more challenging four-source separation task that contains two guitar sources (with no strings). 
    \item \texttt{D0}, \texttt{D1}, \texttt{DF}, and \texttt{DF$_\text{AdaIN}$}: To validate the impact of different choices of spatial information dimension $D$, we investigate two options $D=1$ and $D=F$. Note that \texttt{D0} stands for the XUMX baseline where no spatial conditioning is used, while \texttt{DF$_\text{AdaIN}$} is the $D=F$ case where AdaIN is applied. 
    \item \texttt{CAT} vs. \texttt{ADD}: \texttt{CAT} indicates the combination option that concatenates $|\bX_{:,t}|$ and ${\ba^{(k)}}$. \texttt{ADD}, however, denotes the case when the two are added together. Once again, \texttt{D0} ignores this option.
    \item $\bar\alpha_\text{Tr}$ vs. $\alpha_\text{Tr}$ and $\bar\alpha_\text{Te}$ vs. $\alpha_\text{Te}$: We distinguish the two training cases depending on the type of auxiliary input, i.e., whether the angle is contaminated by the noise ($\bar\alpha_\text{Tr}$) or not ($\alpha_\text{Tr}$). Note that when \texttt{D0}, this training option is turned off and disregarded, as the baseline does not use spatial information. We sample $\epsilon$ from a uniform distribution defined over a range of $[-8, +8]$. There are two types of test experiments defined similarly: $\bar\alpha_\text{Te}$ and $\alpha_\text{Te}$. Our goal is to make sure the system works robustly even on a noisy test signal $\bar\alpha_\text{Te}$. 
\end{itemize}

For example, \texttt{4S2G-D1-CAT-$\alpha_\text{Tr}$-$\bar\alpha_\text{Te}$} indicates a model trained and tested on the two-guitar mixture using the raw source angle added to the spectra as the conditioning mechanism. Here, the auxiliary input is noisy to reflect users' incorrect estimation of the source locations during the test time. However, the model is trained on exact source locations without any noisy angle involved. Meanwhile, \texttt{4S-D0} means the XUMX baseline tested on the default four-source separation experiment with no spatial information involved (or ignored if there is any).

\section{Experimental Results and Discussions}
\label{sec:results}

To evaluate the performance of the various models involved, we consider the following well-established metrics: signal-to-distortion ratio (SDR), source-to-interference ratio (SIR), source-to-artifacts ratio (SAR), and, additionally, source image to spatial distortion ratio (ISR) to properly measure the spatial reconstruction quality in our stereophonic setup \cite{bss_eval_stereo}.

Table~\ref{table:results_task1} presents the results on our first task (\texttt{4S}). We first observe a considerable improvement of at least 11 dB in terms of SDR coming from all of the systems, including our baseline model, over the input mixture. 
More importantly, we note that our first proposed model \texttt{4S-D1-CAT-$\alpha_\text{Tr}$} outperforms the baseline by 1.4 dB on average. Although this scalar raw angle value is imbalanced compared to the high-dimensional spectrum vector, its efficacy signifies the importance of spatial information in MSS.
Furthermore, the proposed positional encoding-based conditioning method successfully brings an additional improvement (0.2 dB) as shown in our \texttt{4S-DF-ADD-$\alpha_\text{Tr}$} models, although the improvement is not too significant. 
Due to the space limitation,  we exclude \texttt{DF-CAT} results that are not too different from \texttt{D1-CAT}, while still being worse.

The injection of noise into $\alpha$ \emph{during training} does not seem to consistently improve the performance if we compare the $\bar\alpha_\text{Tr}$ and $\alpha_\text{Tr}$ models' performance on the noise injected test experiments $\bar\alpha_\text{Te}$. Essentially, it means that the model trained from the accurate source location can still generalize to the test-time inaccurate conditioning. We believe this robustness comes from the fact that (a) the model performs non-spatial source separation anyway (b) the model implicitly extracts and uses the spatial information from the input stereo signal at least to some degree. Meanwhile, \texttt{4S-D1-CAT-$\bar\alpha_\text{Tr}$} does not significantly deteriorate the separation performance on the test set with clean spatial information $\alpha_\text{Te}$.


Table~\ref{table:results_task2} presents the results from our more challenging second task \texttt{4S2G} due to the two overlapping guitar sources that share similar spectral and timbral characteristics. This second task promotes our approach more rightfully as their potentially different spatial positions can dissociate the confusingly overlapping sources. We observe a more substantial improvement from our models over the baseline once again, especially for Gtr1 and Gtr2, of over 3 and 2.6 dB, respectively, on the test signals with accurate angles. The improvement is still substantial when the test source angles are not accurate: 2.6 and 2.4 dB. This jump in performance is also clearly reflected in the SIR scores; particularly for the two guitars where the baseline's SIR is substantially low (-2.3 and -3.2 dB) while our method showcases a clear merit (7.1 and 7.7 dB improvement). This demonstrates how an uninformed system may poorly manage to dissociate between identical instruments while ours may succeed. Once again, due to space-constraints, we opt to exclude \texttt{DF-CAT}, which did not perform nearly as well as \texttt{DF-ADD}. With that in mind, this points us to conclude that the better performance of \texttt{DF-ADD} does not necessarily lie in the size of $D$ but in the conditioning approach. 

In Table \ref{table:results_prelims}  we share additional insight over the different choices of $D$ when they are concatenated to the spectrum as well as the use of the AdaIN option. We found that most of these choices consistently improve the baseline unconditioned model \texttt{D0}, while the simplest \texttt{D1} option shows the best performance.


\section{Conclusions}
\label{sec:conclusion}

In this paper we presented SpaIn-Net, that incorporated a conditioning mechanism for musical source separation, by making use of spatial information. The network was informed of the position of each target source, which could be provided by the user during inference. We proved the benefit of our approach by leading a set of experiments involving diverse musical instrument stems drawn from the Slakh dataset and by exploring different conditioning methods. The outcome of our experiments showed a clear separation improvement and robustness toward incorrect user input on challenging stereo mixtures, both in favor of our method. 
In addition, we showcased a difficult mixing scenario involving multiple instruments of the same class and demonstrated that our approach improved the separation by 2.8 dB on average. While SpaIn-Net showed great promises coupled with a XMUX baseline, we point out that it presents new doors to a relatively unexplored field and that it can serve as a preliminary base for many potential user-centered applications in the future. Source codes and sound examples can be found: \href{https://saige.sice.indiana.edu/research-projects/spain-net}{https://saige.sice.indiana.edu/research-projects/spain-net}

\newpage
\bibliographystyle{IEEEtran}
\bibliography{mjkim}
\end{document}